\documentclass[aps,preprintnumbers,twocolumn,nofootinbib,notitlepage,superscriptaddress,noeprint]{revtex4-2}
\usepackage{natbib}
\usepackage{graphicx}
\usepackage{dcolumn}
\usepackage{bm}
\usepackage{amsmath,amssymb}
\usepackage{float}
\usepackage{multirow}
\usepackage{slashed}
\usepackage{xcolor}
\usepackage{mathtools,braket}
\usepackage{physics}
\usepackage{multirow}
\usepackage[colorlinks=true, pdfstartview=FitV, bookmarks=true, bookmarksnumbered=true, breaklinks]{hyperref}
\usepackage{soul}
\usepackage{lipsum}  
\usepackage{color}
\definecolor{blue}{rgb}{0.0, 0.0, 1.0}
\definecolor{red}{rgb}{1.0, 0.0, 0.0}
\definecolor{royalblue}{rgb}{0.0, 0.14, 0.4}
\hypersetup{linkcolor=blue, citecolor=blue, urlcolor=blue}

\usepackage{hyperref}
\hypersetup{colorlinks=true,citecolor=blue,linkcolor=blue,urlcolor=blue}

\def\orcid#1{\kern .08em\href{https://orcid.org/#1}
{\includegraphics[keepaspectratio,width=0.7em]{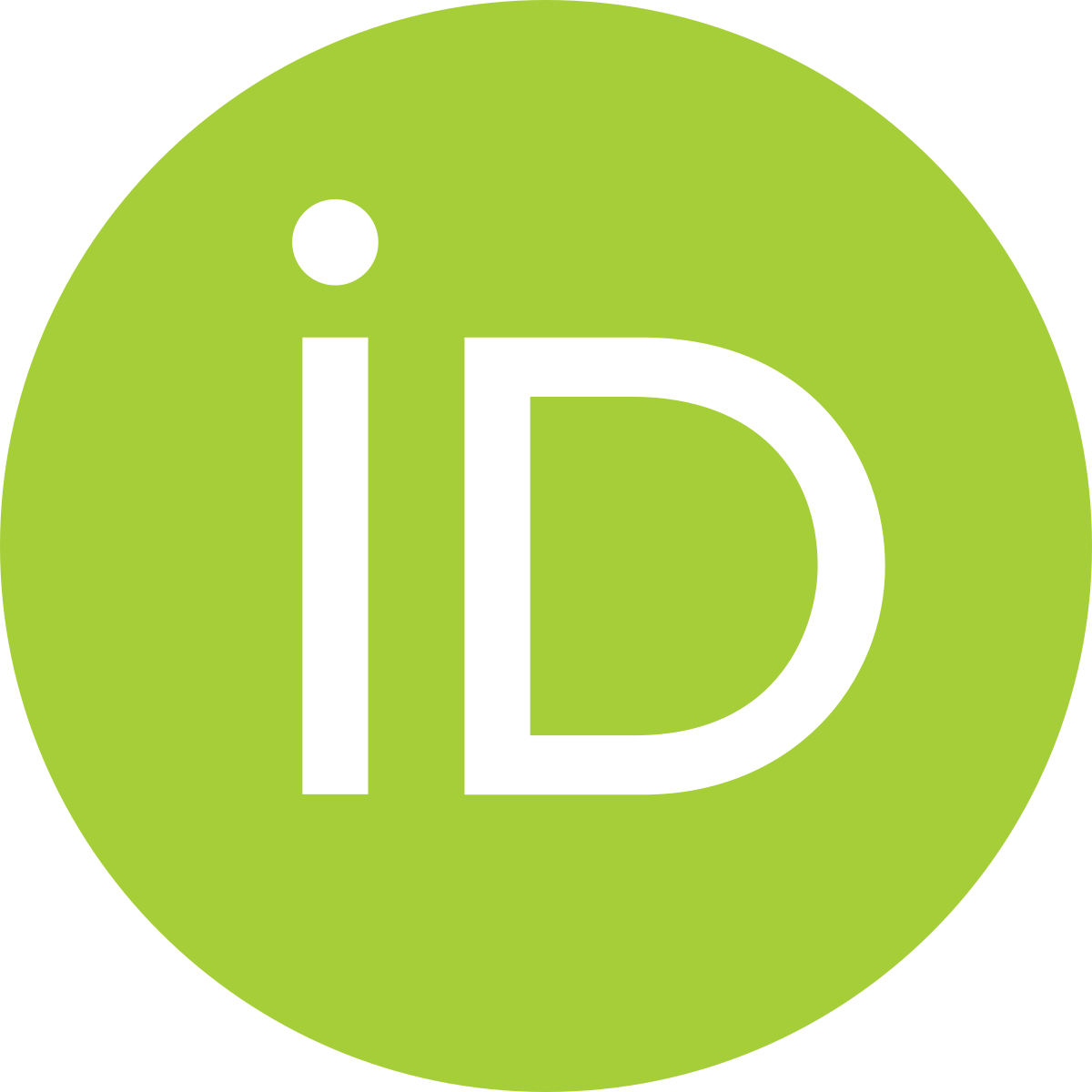}}}
\usepackage[mathlines]{lineno}

\usepackage[normalem]{ulem}
\usepackage{color}

\begin{document}
%\vspace{-0.5cm}
%  \begin{flushright}
%    {\bf LFTC-24-11/94}
%  \end{flushright}
%\vspace{0.5cm}
%----------------------------------------------------------------
\preprint{LFTC-24-11/94}
%----------------------------------------------------------------
\title{In-medium electromagnetic form factors of pseudoscalar mesons from the quark model}
%----------------------------------------------------------------

%----------------------------------------------------------------
\author{Ahmad Jafar Arifi\orcid{0000-0002-9530-8993}}
\email{ahmad.arifi@riken.jp} 
\affiliation{Few-Body Systems in Physics Laboratory, RIKEN Nishina Center, Wako 351-0198, Japan}
\affiliation{Research Center for Nuclear Physics (RCNP), The University of Osaka, Ibaraki 567-0047, Japan}
%----------------------------------------------------------------
\author{Parada~T.~P.~Hutauruk\orcid{0000-0002-4225-7109}}
\email{phutauruk@pknu.ac.kr} 
\affiliation{Department of Physics, Pukyong National University (PKNU), Busan 48513, South Korea}
%----------------------------------------------------------------
\author{Kazuo Tsushima\orcid{0000-0003-4926-1829}}
\email{kazuo.tsushima@cruzeirodosul.edu.br; kazuo.tsushima@gmail.com} % 
\affiliation{Laboratório de Física Teórica e Computacional-LFTC, 
Programa de P\'{o}sgradua\c{c}\~{a}o em Astrof\'{i}sica e F\'{i}sica Computacional, Universidade Cidade de S\~{a}o Paulo, 01506-000 S\~{a}o Paulo, SP, Brazil}
%----------------------------------------------------------------

\date{\today}

%----------------------------------------------------------------
\begin{abstract}
We explore the modifications of hadron structure in a nuclear medium, focusing on the spacelike electromagnetic form factors (EMFFs) of light and heavy-light pseudoscalar mesons. 
By combining the light-front quark model (LFQM) with the quark-meson coupling (QMC) model, which reasonably reproduces EMFFs in free space and the saturation properties of nuclear matter, respectively, we systematically analyze the in-medium EMFFs and charge radii of mesons with various quark flavors.
Our findings show that the EMFFs of charged (neutral) mesons exhibit a faster fall-off (increase) with increasing four-momentum transfer squared and nuclear density. 
Consequently, the absolute value of the charge radii of mesons increases with nuclear density, where the rate of increase depends on their quark flavor contents.
We observe that the EMFFs of pions and kaons undergo significant modifications in the nuclear medium, while heavy-light mesons are only slightly modified.
By decomposing the quark flavor contributions to EMFFs, we show that the medium effects primarily impact the light-quark sector, leaving the heavy-quark sector nearly unaffected. The results of this study further suggest the importance of the medium effects at the quark level.
\end{abstract}
%----------------------------------------------------------------

\maketitle

%================================================================
\section{Introduction}
%================================================================

Modifications of the hadron properties and structure in a nuclear medium and nuclei have been explored through a variety of perspectives. They can provide us with important and useful insights into the complicated hadron dynamics in the nuclear medium~\cite{Hayano:2008vn,Hosaka:2016ypm,Leupold:2009kz,Post:2003hu,Metag:2017yuh,Tolos:2020aln}.
One of the broadly known examples is the modifications of the structure functions of the bound hadrons in nuclei compared with those in the free space, where the phenomenon is well known as the European Muon Collaboration (EMC) effect~\cite{Aubert:1983wih}. 
This effect highlights the influence of the nuclear medium on the internal quark-gluon dynamics of hadrons, suggesting modifications to their quark-level structure, and driving advanced developments and improvements in nuclear matter models based on the quark-level dynamics~\cite{Guichon:2018uew,Li:2020dst,Bentz:2001vc, Mineo:2003vc,Fuchs:1995as,Fornetti:2023gvf}.
Moreover, measuring the EMC effect further in detail is one of the main Electron-Ion Collider (EIC) scientific programs~\cite{Accardi:2012qut}. 
The measurements are expected to shed light on the detailed mechanisms of the nuclear modifications and will provide critical tests for the theoretical predictions and models.

In addition to the EMC effect, hadrons in the nuclear medium undergo various modifications~\cite{Montesinos:2024xpp,KEK-PS-E325:2005wbm,JeffersonLabE93-049:2002asn,Suzuki:2002ae}, such as (effective) mass modifications, width broadening, and the increase of their charge radii. 
Mass modifications in the nuclear medium that are associated with the changes in the scalar mean field in some models, which might be expected to link to the quark chiral condensate, 
signaling partial restoration of chiral symmetry, 
while the width-broadening modifications from enhanced interaction rates and modified decay channels in the dense nuclear medium are due to the competition between the scalar and vector meson mean fields in usual relativistic mean field models. Increases in charge radii of hadrons indicate the medium modifications in the electromagnetic form factors (EMFFs), as the charge distribution becomes more diffused due to medium effects. These phenomena in a nuclear medium may be connected to chiral symmetry restoration, and this idea is supported by experimental evidence of deeply bound pionic~\cite{Suzuki:2002ae} and kaonic~\cite{Sgaramella:2024qdh} atoms, and by the associated various theoretical studies~\cite{Itahashi:2024apl,Gifari:2024ssz,Hutauruk:2018qku,Hutauruk:2019was,Jido:2008bk}. Thus, they will provide new insights into the interplay between the hadron structure and the effects of nuclear medium. 

Our main focus in this study is on the nuclear medium modifications of the meson spacelike EMFFs. 
In free space, the EMFFs of the mesons have been extensively studied using various theoretical models, e.g. in Refs.~\cite{Bijnens:2002hp,Choi:1997iq,Arifi:2024mff,Moita:2021xcd,Hutauruk:2016sug,Yao:2024drm,Abidin:2019xwu,Burden:1995ve,Chang:2013nia,Maris:1998hc,Braguta:2007fj,Maris:2000sk}, including the chiral-perturbation theory~\cite{Bijnens:2002hp}, the light-front quark models~\cite{Choi:1997iq,Arifi:2024mff,Moita:2021xcd}, 
the Bethe--Salpeter--Equations Nambu--Jona--Lasinio (BSE NJL) 
model~\cite{Hutauruk:2016sug}, the Holographic QCD or AdS/CFT correspondence model~\cite{Abidin:2019xwu}, the Dyson-Schwinger Equation (DSE) based approach~\cite{Burden:1995ve,Chang:2013nia} among others. 
These model results generally provide reasonable explanations for the available data on EMFFs in the low-energy region.
Furthermore, investigations of the EMFFs of the pion and kaon at high $Q^2$ up to approximately $6\, \text{GeV}^2$ are currently being conducted as part of the ongoing JLab 12 GeV program~\cite{Horn:2019,Arrington:2021alx}, with further measurements planned over a wider range of $Q^2$ at the future EIC facility~\cite{Accardi:2012qut}.
A number of lattice QCD simulations, \textit{e.g.} in Refs.~\cite{QCDSF:2017ssq,Wang:2020nbf,Can:2012tx,Li:2017eic,Koponen:2015tkr,Aoki:2015pba,Alexandrou:2021ztx,Ding:2024lfj,Gao:2021xsm}, have also been performed to study EMFFs in various contexts.

In contrast, studies on the EMFFs of the light and heavy-light pseudoscalar mesons in the nuclear medium are limited so far.
Since lattice QCD simulations are not yet established at lower and higher nuclear densities due to the sign problem~\cite{Muroya:2003qs}, 
most studies rely on effective model calculations.
For instance, the EMFFs of the pion and kaon have been investigated in the nuclear medium 
using the NJL model in combination with the quark-meson coupling (QMC) 
model~\cite{Hutauruk:2018qku, Hutauruk:2019was,Hutauruk:2019ipp}. 
Very recently, a consistent calculation and treatment on hadron structure and nuclear matter 
using the same NJL model was reported in Ref.~\cite{Hutauruk:2021kej,Gifari:2024ssz}. 
Additionally, the EMFFs of the pion~\cite{deMelo:2014gea} and kaon~\cite{Yabusaki:2023zin} have been studied in the light-front quark model (LFQM) combined with the QMC model with the help of the Bethe-Salpeter approach (BSA), as well as using QCD sum rules~\cite{Er:2022cxx}. 
Most of the studies focus on the EMFFs of the pion and kaon, and it has been limited
for studying the EMFFs of the heavy-light mesons.
In addition to our focus of study, it is worth mentioning that the EMFFs of the octet baryons in the nuclear medium have also been made in the covariant spectator quark model~\cite{Ramalho:2012pu,Ramalho:2019ues}, the QMC model~\cite{Guichon:1995ue} and LFQM~\cite{deAraujo:2017uad}.
Notably, the double ratio of the nucleon EMFFs in nuclei can be measured rather accurately in experiments~\cite{JeffersonLabE93-049:2002asn}. This shows that studying the medium effects on hadron structure deserves further investigation.

In our previous study~\cite{Arifi:2023jfe}, we have developed  
an approach combined the LFQM with the QMC model to compute the in-medium decay constants and distribution amplitudes (DAs) of the light and heavy-light mesons, where such studies have also been made in other approaches~\cite{Puhan:2024xdq, deMelo:2016uwj, Hutauruk:2019ipp,Hutauruk:2021kej,Bozkir:2022lyk}. Based on our previous approach, we now extend to study the in-medium modifications of the EMFFs and charge radii of the light and heavy-light pseudoscalar mesons in the symmetric nuclear matter (SNM). The present study provides a systematic analysis of the in-medium EMFFs of the mesons with different quark flavor contents, $u,d,s,c$ and $b$. 

This article is organized as follows. In Sec.~\ref{sec:LFQM}, we briefly describe the computation of the light front wave functions (LFWFs) in the framework of the LFQM and its normalization.  Section~\ref{sec:QMC} presents the properties of quarks and the light and heavy-light pseudoscalar mesons in SNM computed in the QMC model. In  Sec.~\ref{sec:eff}, we use these in-medium quark and meson properties and calculate the EMFFs and charge radii of the light and heavy-light pseudoscalar mesons. Numerical results on the EMFFs and charge radii of the light and heavy-light pseudoscalar mesons in SNM are presented in Sec.~\ref{sec:num}. 
Finally, a summary is given in Sec.~\ref{sec:sum}.

%================================================================
\section{Model description}
%================================================================

In Sec.~\ref{sec:LFQM}, we provide a brief overview of the free-space properties of the light and heavy-light pseudoscalar mesons using the LFQM, which is built on the constituent quark model with light-front dynamics. Following this, Sec.~\ref{sec:QMC} presents a summary of the QMC model, where the properties of the effective quarks and mesons in the nuclear medium are calculated.

%================================================================
\subsection{Light-front quark model} \label{sec:LFQM}
%================================================================

Here, we begin with outlining the main concepts of the LFWF construction within the LFQM~\cite{Arifi:2023jfe}. 
It is important to emphasize that our approach differs from 
the Bethe–Salpeter amplitude-based method used in Refs.~\cite{deMelo:2014gea,Yabusaki:2023zin}. 
Specifically, our approach utilizes a vertex function regulated by a Gaussian form and the meson states are built using the Bakamjian–Thomas (BT) construction~\cite{Bakamjian:1953kh,Keister:1991sb}, 
which maintains Poincar\'{e} invariance and ensures independence from any particular frame.
Self-consistent studies of several observables~\cite{Ridwan:2024ngc,Arifi:2023uqc,Arifi:2022qnd,Choi:2024ptc} are made using the BT construction.

In the LFQM, the meson state $\ket{{\cal M}} \equiv \ket{{\cal M} (P,J, J_z)}$, as a bound state of the constituent 
quark $q$ and antiquark $\Bar{q}$ with meson momentum $P$ and total angular momentum $(J,J_z)$, 
can be written as
%%%%%%%%%5
\begin{eqnarray}
\ket{{\cal M}} 
&=& \int \left[ \dd ^3\bm{p}_q \right] \left[ \dd ^3\bm{p}_{\bar{q}} \right]  2(2\pi)^3 \delta^3 
\left(\bm{P}-\bm{
p}_q - \bm{p}_{\bar{q}} \right) 
\nonumber\\ && \times \mbox{} 
\sum_{\lambda_q,\lambda_{\bar{q}}} \Psi_{\lambda_q \lambda_{\bar{q}}}^{JJ_z}(x, \bm{k}_\perp)
\ket{q_{\lambda_q}(p_q) \bar{q}_{\lambda_{\bar{q}}}(p_{\bar{q}}) },
\quad\quad 
\end{eqnarray}
where $\bm{p_i}=(p_i^+,\bm{p}_{i \perp})$ and $\left[ \dd ^3\bm{p}_i \right] \equiv {\rm 
d}p_i^+ \dd ^2\bm{p}_{i \perp}/[2(2\pi)^3]$. 
We represent the momenta and helicities of the quark and antiquark as $(p_q,\lambda_q)$ and $(p_{\bar{q}},\lambda_{\bar{q}})$ for $i=q$ and $i={\bar{q}}$, respectively. The internal light-front variables $(x, \bm{k}_\perp)$ are then defined by $x = p^+_q/P^+$ and $\bm{k}_{\perp} = x \bm{P}_\perp - \bm{p}_{q\perp}$.

The LFWF of the ground state pseudoscalar meson in momentum space is given by
%%%%%%%%%%
\begin{eqnarray}\label{eq:5}
		\Psi^{JJ_z}_{\lambda_q\lambda_{\bar{q}} }(x, \bm{k}_{\bot}) = \Phi(x, \bm{k}_\bot)
		\  \mathcal{R}^{JJ_z}_{\lambda_q\lambda_{\bar{q}} }(x, \bm{k}_\bot),
\end{eqnarray}
%%%%%
where $\Phi(x, \bm{k}_\bot)$ and $\mathcal{R}^{JJ_z}_{\lambda_q\lambda_{\bar{q}}}(x, \bm{k}_\bot)$ represent the radial and spin-orbit wave functions, respectively.
The matrix element $\mathcal{R}^{JJ_z}_{\lambda_q\lambda_{\bar{q}}}$ is derived through the Melosh transformation~\cite{Melosh:1974cu} and can be expressed in covariant form as
%%%%
\begin{eqnarray}\label{eq:6}
	\mathcal{R}^{00}_{\lambda_q\lambda_{\bar{q}}} &=&  \frac{1}{\sqrt{2} \tilde{M}_0} 
	\bar{u}_{\lambda_q}^{}(p_q) \gamma_5 v_{\lambda_{\bar{q}}}(p_{\bar{q}}),
\label{M0til}	
\end{eqnarray}
%%%%
with $\tilde{M}_0 \equiv \sqrt{M_0^2 - (m_q -m_{\bar{q}})^2}$,
and the invariant meson mass $M_0^2$ is defined as
%%%%%%
\begin{eqnarray}\label{eq:4}
	M_0^2 = \frac{\bm{k}_{\bot}^2 + m_q^2}{x}  + \frac{\bm{k}_{\bot}^2 + m_{\bar{q}}^2}{1-x}.
\end{eqnarray} 
%%%%%% 
The explicit form of the spin-orbit wave functions for the pseudoscalar is given by
\begin{equation}
\mathcal{R}^{00}_{\lambda_q\lambda_{\bar{q}}} (x,\bm{k}_\perp) 
=\frac{1}{\sqrt{2}\sqrt{\mathcal{A}^2 + \bm{k}_\perp^2}}
\begin{pmatrix}
k^L 		& \mathcal{A}\\
-\mathcal{A} & k^R\\
\end{pmatrix},
\end{equation}
with $k^{R(L)}=k_x\pm i k_y$ and $\mathcal{A}=xm_{\bar{q}} + (1-x) m_q$. It is worth noting that the spin-orbit wave functions satisfy the orthonormality condition, expressed as
%#######################################
\begin{eqnarray}
\sum_{\lambda_q,\lambda_{\bar{q}}} \braket{\mathcal{R}^{JJ_z}_{\lambda_q\lambda_{\bar{q}}} }{\mathcal{R}^{J^\prime J_z^\prime}_{\lambda_q\lambda_{\bar{q}}}} = \delta_{JJ^\prime}\delta_{J_z J_z^\prime}.
\end{eqnarray}
%#######################################

For the meson ground state, we employ a phenomenological Gaussian wave function as 
%%%%%%
\begin{eqnarray}\label{eq:9}
	\Phi_{1S} (x, \bm{k}_\bot) &=& \frac{4\pi^{3/4}}{ \beta^{3/2}} 
	\sqrt{\frac{\partial k_z}{\partial x}} e^{-\bm{k}^2/ 2\beta^2},
\end{eqnarray}
%%%%%
where $\beta$ is the variational parameter associated with the wave function's size, and the Jacobian factor is represented by
%%%%%%
\begin{equation}
\frac{\partial k_z}{\partial x} = \frac{M_0}{4x(1-x)} \left[ 1 - \frac{ (m_q^2 - m_{\bar{q}}^2)^2}{M_0^4} 
\right],
\end{equation}
%%%%%
which accounts for the transformation of variables from $(k_z,\bm{k}_\perp)$ to $(x,\bm{k}_\perp)$, where 
\begin{eqnarray}
    k_z = \left( x - \frac{1}{2} \right) M_0 + \frac{(m^2_{\bar{q}} -m^2_q)}{2M_0}.
\end{eqnarray} 
The LFWF is subsequently normalized by
%%%%%
\begin{eqnarray}\label{eq:10}
 \int \frac{\dd x \dd ^2 \bm{k}_\bot}{2(2\pi)^3}  \abs{ \Psi (x, \bm{k}_\bot) }^2 =1.
\end{eqnarray}
%%%%%

%================================================================
\subsection{Quark-meson coupling model} \label{sec:QMC}
%================================================================

Before examining the effects of the in-medium modifications of the light and heavy-light meson structure, we briefly overview the QMC model in this section. In this model, parameters are determined to reproduce the saturation properties of the nuclear matter. The QMC model, which describes the nuclear matter, (hyper)nuclei, and hadron properties in the nuclear medium based on the quarks utilizes the relativistic mean field approximation~\cite{Saito:2005rv,Guichon:1987jp,Saito:1996sf,Guichon:2018uew,Krein:2017usp}.
In this framework, the meson mean fields interact directly with the confined light $u$ and $d$ valence quarks constructed by the MIT bag model.

%----------------------------------------------------------------
\subsubsection{Relativistic mean-field approximation}
%----------------------------------------------------------------
The effective Lagrangian density for the symmetric nuclear matter at the hadronic level is given by~\cite{Guichon:1987jp,Saito:2005rv,Saito:1996sf,Guichon:2018uew} 
%%%%%%%%
\begin{eqnarray}
\mathcal{L}_{\mathrm{QMC}}  &=& \mathcal{L}_{\rm nucleon} + \mathcal{L}_{\rm meson} + 
\mathcal{L}_{\rm int},
\end{eqnarray}
with each component defined as follows
%%%%%%%%
\begin{eqnarray}
\mathcal{L}_{\rm nucleon} &=& 
 \bar{\psi}[i\slashed{\partial} - m_N ]\psi, 
\\
\mathcal{L}_{\rm meson} &=& \frac{1}{2} ( \partial_\mu \hat{\sigma}\partial^\mu\hat{\sigma} - 
m_\sigma^2\hat{\sigma}^2  ) \nonumber\\
& & - \frac{1}{2} \left[ \partial_\mu \hat{\omega}_\nu (\partial^\mu \hat{\omega}^\nu - 
\partial^\nu 
\hat{\omega}^\mu ) - m_\omega^2 \hat{\omega}^\mu \hat{\omega}_\mu\right],\quad\quad 
\end{eqnarray}
where $\psi$, $\hat{\sigma}$, and $\hat{\omega}$ denote the field operators for the nucleon, $\sigma$, and $\omega$ fields, respectively. 
The corresponding interaction Lagrangian density is expressed as
%%%%%%%
\begin{eqnarray}
\mathcal{L}_{\rm int} =\tilde{g}_{N\sigma}(\hat{\sigma}) \bar{\psi} \psi \hat{\sigma}  - g_{N\omega} 
\hat{\omega}^\mu 
\bar{\psi} \gamma_\mu \psi,
\end{eqnarray} 
where $\tilde{g}_{N\sigma}(\hat{\sigma})$ is the $\sigma$-field dependent $N\sigma$ coupling constant, and $g_{N\omega}$ represents the $N\omega$ coupling constant.

The Lagrangian density can alternatively be written as
\begin{eqnarray}
\mathcal{L}  &=&  \bar{\psi}[i\slashed{\partial} - m_N^*(\hat{\sigma}) - g_{N\omega} \hat{\omega}^\mu \gamma_\mu  ]\psi + \mathcal{L}_{\rm meson},
\end{eqnarray}
where the nucleon's effective mass at a given density is defined by
%%%%%%%%
\begin{eqnarray}
\label{eqmNmed}
m_N^*(\hat{\sigma}) = m_N - \tilde{g}_{N\sigma}(\hat{\sigma}) \hat{\sigma}. 
\end{eqnarray} 
The coupling $\tilde{g}_{N\sigma}(\hat{\sigma})$ influences the nucleon effective mass, modifying the nucleon mass in a nonlinear way through the $\sigma$ field. 
On the other hand, the $g_{N\omega}$ coupling modifies the nucleon’s four-momentum.
In the mean-field approximation, the meson field operators are replaced by their constant mean field expectation values, i.e., $\hat{\sigma} \to \sigma = \langle \hat{\sigma} \rangle$ and $\hat{\omega}_\mu \to \delta_{\mu,0}\, \omega = \langle \hat{\omega}_\mu \rangle$.

At the nucleon level, the equations of motion of the meson fields are given by
%%%%%%
\begin{eqnarray} \label{eq:eom}
(\Box + m_\sigma^2)\sigma &=& \left(-\frac{\partial m_N^*(\sigma)}{\partial \sigma}\right) 
(\bar{\psi}\psi)  = 
\tilde{g}_{N\sigma}(\sigma) \rho_s, \quad\quad  \\
(\Box + m_\omega^2)\omega &=& g_{N\omega} (\bar{\psi}\gamma^0\psi) = g_{N\omega} (\psi^\dagger\psi)  
=g_{N\omega} \rho,
\label{eq:eom2}
\end{eqnarray}
where in nuclear matter, the d'Alembert operator $\Box$ is set to zero, and $\rho_s$ and $\rho$ represent the nucleon scalar and vector (baryon) densities, respectively.

The Dirac equation for the nucleon is given by
%%%%%%%
\begin{eqnarray}
(i\slashed{\partial} - g_{N\omega} \omega \gamma^0 - m_N^{*}(\sigma) )\psi = 0.
\end{eqnarray}
Here, the effective nucleon mass appears in this equation and in Eq.~(\ref{eq:eom}) as
%%%%%%
\begin{eqnarray}
- \frac{\partial  m_N^*(\sigma) }{\partial \sigma} &=& \tilde{g}_{N\sigma}(\sigma) = g_{N\sigma} 
C_N(\sigma), 
\end{eqnarray}
where
%%%%%%%%
\begin{eqnarray}
C_N(\sigma) = \frac{S_N(\sigma)}{S_N(\sigma=0)}. 
\end{eqnarray}
The function $C_N(\sigma)$ represents the scalar polarizability, which characterizes the nucleon’s response to the external scalar field~\cite{Saito:2005rv}.
For a point-like nucleon, $C_N(\sigma)=1$.

In the QMC model, the meson-nucleon couplings $(g_{N\sigma}, g_{N\omega})$ are related to the quark-meson couplings $(g_{q\sigma}, g_{q\omega})$ as follows
%%%%%%
\begin{eqnarray}
g_{N\sigma} &=& \tilde{g}_{q\sigma}(\sigma=0) = 3g_{q\sigma} S_N(\sigma=0),\\
g_{N\omega} &=& 3g_{q\omega},
\label{gNs}
\end{eqnarray}
where $S_N(\sigma) = \int \dd^3x \bar{\psi}_q(x) \psi_q(x)$ is calculated using the MIT bag model as given in Eq.~\eqref{eq:polarizability}. The factor of three reflects the fact that the nucleon consists of three light valence quarks.

From Eqs.~(\ref{eq:eom}) and~(\ref{eq:eom2}), the vector and scalar meson fields are calculated as follows
%%%%%%%
\begin{eqnarray}
\omega = \frac{g_{N\omega} \rho}{m_\omega^2}, \qquad 
\sigma = \frac{g_{N\sigma} \rho_s}{m_\sigma^2} C_N(\sigma), \label{eq:meanfield}
\end{eqnarray}
where the nuclear density $\rho$ and scalar density $\rho_s$ are given by
%%%%%%%
\begin{eqnarray}
\rho &=&  \frac{4}{(2\pi)^3} \int \dd^3\bm{k}\ \Theta(k_F - k) = \frac{2k_F^3}{3\pi^2}, 
\label{eq:kf}\\
\rho_s &=& \frac{4}{(2\pi)^3} \int \dd^3\bm{k}\ \Theta(k_F - k) 
\frac{m_N^*(\sigma)}{\sqrt{m_N^{*2}(\sigma) + 
k^2}},
\end{eqnarray}
where $k = |\bm{k}|$ and $\Theta(k_F - k)$ is the step function ensuring the integral is performed up to the nucleon Fermi momentum $k_F$, which is related to the nuclear density $\rho$. The factor of four accounts for spin and isospin degeneracy. 
As shown in Eq.~(\ref{eq:meanfield}), we solve the self-consistent equation for the $\sigma$ mean-field to determine its value at each nuclear density.

Once the $\sigma$ and $\omega$ mean fields are obtained, the total energy per nucleon can be computed as
%%%%%%%
\begin{eqnarray}
\frac{E_{\rm tot}}{A} &=& \frac{1}{\rho} \biggl[ \frac{4}{(2\pi)^3}\int \dd^3\bm{k}\ \Theta(k_F - 
k) 
\sqrt{m_N^{*2}(\sigma) +k^2 } \nonumber\\
 & & + \frac{1}{2} g_{N\sigma} C_N(\sigma) \sigma \rho_s  + \frac{1}{2}g_{N\omega} \omega \rho 
\biggr].
\end{eqnarray}
The model parameters are then determined by fitting the nuclear matter saturation properties at the saturation density $\rho_0=0.15$ fm$^{-3}$ ($k_F=1.305$ fm$^{-1}$), such as the negative of the binding energy ($E_{\rm tot}/A - m_N$), which is $-15.7$ MeV.

%----------------------------------------------------------------
\subsubsection{MIT bag model}
%----------------------------------------------------------------

In the standard QMC model~\cite{Guichon:1987jp, Guichon:1995ue, Saito:1996sf, Guichon:2018uew}, the nucleon-meson couplings are derived from the quark-meson couplings. This is done by using the MIT Bag model for hadrons, and solving the Dirac equations for the quarks and antiquarks in the presence of meson mean fields in nuclear matter. The meson potentials are given by
\begin{eqnarray}
V_{q\sigma} = g_{q\sigma} \sigma, \qquad V_{q\omega} = g_{q\omega} \omega.
\end{eqnarray}
Here, $q$ refers to light quarks ($u$ or $d$), and $Q$ represents heavier quarks ($s, c, b$) confined within the bag of a hadron in SNM, with $r = |\bm{r}|$ up to the bag radius.

The Dirac equations for the quark and antiquark in the presence of the mean-field potentials are given by
\begin{eqnarray}
\left[i\slashed{\partial} - (m_q - V_{q\sigma}) \mp \gamma^0 V_{q\omega}\right]
\begin{pmatrix}
\psi_q(z) \\
\psi_{\bar{q}}(z) \\
\end{pmatrix} = 0, \\
\left[i\slashed{\partial} - m_Q \right]
\begin{pmatrix}
\psi_Q(z) \\
\psi_{\bar{Q}}(z) \\
\end{pmatrix} = 0,
\end{eqnarray}
where we assume $m_q = m_u = m_d$ for the light quarks under SU(2) symmetry. The scalar potential modifies the quark mass, while the vector potential shifts the quark energy in the nuclear medium. The effective quark mass is given by
\begin{eqnarray}
m_q^* = m_q - V_{q\sigma}.
\end{eqnarray}
It is important to note that the mean fields couple only to the light quarks and antiquarks. For heavier quarks ($Q = s, c, b$), their masses remain the same in the nuclear medium as in free space ($m_Q^* = m_Q$), since the $\sigma$ field does not couple to these heavier quarks. 

We derive the static solution for the ground state quark or antiquark, where the Hamiltonian is time-independent, and the wave function is written as
\begin{eqnarray}
\psi(z) = \psi(r) \exp\left(-i\varepsilon^* t/R^*\right),
\end{eqnarray}
with $R^*$ being the in-medium bag radius. The eigenenergies in units of $1/R^*$ are
\begin{eqnarray}
\begin{pmatrix}
 \varepsilon^*_q \\
\varepsilon^*_{\bar{q}} \\
\end{pmatrix} &=& \Omega_q^* \pm R^* V_\omega^q,
\end{eqnarray}
where 
\begin{eqnarray}
    \Omega^*_q &=& \sqrt{x_q^{*2} + (m_q^* R^*)^2},
\end{eqnarray}
with $x_q^*$ representing the lowest mode bag eigenvalue. The normalized ground state quark eigenfunction is
\begin{eqnarray}
\psi(z) = \frac{N \, {\rm e}^{-i\varepsilon^* t/R^*}}{\sqrt{4\pi}} \begin{pmatrix}
    j_{0}(x_q^* r/R^*) \\ i\beta_q^* j_{1}(x_q^* r/R^*)\ {\bm \sigma} \cdot \hat{r} 
\end{pmatrix}  \chi_m,
\end{eqnarray}
where $\chi_m$ is the spin function and $j_{0,1}(r)$ are spherical Bessel functions. The normalization constant $N$ is determined from
\begin{eqnarray}
\int_0^{R^*} \dd ^3\bm{r}\  \psi^\dagger(r) \psi(r) = 1.
\end{eqnarray}
This gives
\begin{eqnarray}
N^{-2} = 2 R^{*3} j_0^2(x_q^*) \frac{[\Omega_q^*(\Omega_q^*-1) + m_q^* R^*/2]}{x_q^{*2}}.
\end{eqnarray}
The eigenfrequency of the quark is obtained by ensuring the continuity of the quark eigenfunction at the bag boundary ($r=R^*$), leading to the relation
\begin{eqnarray}
j_0(x_q^*) = \beta_q^* j_1(x_q^*),
\end{eqnarray}
with 
\begin{eqnarray}
\beta_q^* = \sqrt{\frac{\Omega_q^* - m_q^* R^*}{\Omega_q^* + m_q^* R^*}}.
\end{eqnarray}

By solving the equation above, we find the lowest positive eigenvalue of $x_q^*$. The effective nucleon mass in the MIT bag model is given by
\begin{eqnarray}
m_N^*(\sigma) = \frac{3\Omega^*_q - Z_N}{R^*} + \frac{4\pi R^{*3}}{3} B,
\end{eqnarray}
where $Z_N$ accounts for gluon fluctuation and center-of-mass motion corrections~\cite{Guichon:1995ue}, and $B$ is the bag pressure. 
The equilibrium is reached when the nucleon mass is minimized, leading to
\begin{eqnarray}
\left.\frac{\dd  m_N^*(R^*)}{\dd  R^*}\right|_{R^*=R_N^*} = 0,
\end{eqnarray}
which determines the nucleon radius $R_N^*$ and its mass self-consistently with $x_q^*$. 

It is important to note that the bag radius $R_N^*$ is not a physical observable unlike the nucleon radius, which requires computing from the quark wave function. Additionally, the scalar polarizability $C_N(\sigma)$ is related to the $\sigma$-dependent $N\sigma$ coupling constant $\tilde{g}_\sigma^N(\sigma)$, with
\begin{eqnarray}\label{eq:polarizability}
S_N(\sigma) = \frac{\Omega_q^*/2 + m_q^* R^*(\Omega_q^*-1)}{\Omega_q^*(\Omega_q^*-1) + m_q^* R^*/2}.
\end{eqnarray}

This approach offers a new perspective for the saturation properties of nuclear matter, based on the quark structure of the nucleon. It eliminates the need for nonlinear meson field couplings in the effective Lagrangian to achieve a reasonable incompressibility value of $K \simeq 200 - 300$ MeV, commonly practiced in relativistic mean-field models~\cite{Serot:1997xg}.

%================================================================
\section{Electromagnetic form factors} \label{sec:eff}
%================================================================

In this section, we first discuss the computation of the EMFFs in free space by using the LFQM. Subsequently, we introduce the in-medium effect to study the modified EMFFs.

%================================================================
\subsection{EMFFs in free space}
%================================================================

In free space, the EMFFs of the pseudoscalar mesons $F_M(q^2)$ are defined in the matrix element as
%%%%%%%%%%%%%%%%
\begin{eqnarray}
\label{eq:ffm}
    \left< P^\prime \right| j^\mu_{em}\left| P\right> = (P^{\prime\mu} + P^\mu) F_M(q^2),
\end{eqnarray}
%%%%%%%%%%%%%%%
where $q^2=(P^\prime -P)^2$ is the four-momentum transfer squared. 
To compute EMFFs, we use the Drell-Yan-West frame ($q^+=0$), the space-like momentum transfer $\bm{q}_\perp^2\equiv Q^2 = -q^2$ where $\bm{P}_\perp=0$.
From the quark momentum conservation, we obtain $\bm{k}_\perp^\prime = \bm{k}_\perp +(1-x) \bm{q}_\perp.$

In LFQM~\cite{Choi:1997iq,Arifi:2022pal}, the plus current component ($\mu=+$) is typically used to compute the EMFFs due to its simplicity since $(P^{\prime+} + P^+)=2P^+$ is merely a kinematical factor.
Alternatively, the transverse ($\mu=\perp$) and minus ($\mu=-$) current components can also be employed; 
although achieving self-consistency in such cases may require modifications to the Lorentz structure on the right-hand side of Eq.~\eqref{eq:ffm} in the BT construction, as discussed in Ref.~\cite{Choi:2024ptc}.

The EMFFs of the pseudoscalar mesons can be decomposed by their quark sector form factors as
%%%%%%%%%%%%%%%%
\begin{eqnarray}
    F_M(Q^2) = e_q F^q_{M}(Q^2, m_q, m_{\bar{q}}) + e_{\Bar{q}} F^{\bar{q}}_{M}(Q^2, m_{\bar{q}}, m_q), \quad 
\end{eqnarray}
%%%%%%%%%%%%%%%
where $e_q (e_{\Bar{q}})$ is the electric charge of the quark (antiquark).
The quark sector form factor can be computed within the impulse approximation, where they are obtained through a convolution of the initial and final LFWFs as
\begin{eqnarray}
F^{q}_M &=& \int \frac{\dd x\ \dd^2 \bm{k}_\bot}{16\pi^3}\ \frac{  \Phi(x,\bm{k}_\perp) \Phi'(x,\bm{k}^\prime_\perp) }{2P^+}\nonumber\\
& & \sum_{\lambda,\lambda^\prime,\bar{\lambda}} \mathcal{R}_{\lambda^\prime \bar{\lambda}}^{00\dagger}(x, \bm{k}_\perp^\prime) \frac{\bar{u}_{\lambda^\prime}(p_1^\prime) }{\sqrt{x}} \gamma^{\mu} 
\frac{u_{\lambda}(p_1) }{\sqrt{x}} \mathcal{R}_{\lambda \bar{\lambda}}^{00}(x, \bm{k}_\perp),\nonumber\\
\end{eqnarray}
where only the helicity non-flip process contribute to the EMFFs when using plus current component since 
$\bar{u}_{\lambda^\prime}(p_1^\prime) \gamma^+ u_{\lambda}(p_1) = 2\sqrt{p_1^+p_1^{\prime +}}\delta_{\lambda\lambda^\prime}.$
The quark sector form factors are explicitly expressed as~\cite{Choi:1997iq}
%%%%%%%%%%%%%%%
\begin{eqnarray}
\label{eq:emffs}
    F^{i}_{M}(Q^2) = \int_0^1 \dd x \int \frac{\dd ^2\bm{k}_\perp}{2(2\pi)^3} \Phi(x,\bm{k}_\perp) \Phi^\prime(x,\bm{k}_\perp^\prime) \nonumber\\ 
     \times \frac{\bm{k}_\perp\cdot\bm{k}_\perp^\prime + \mathcal{A}^2}{\sqrt{\bm{k}_\perp^2 + \mathcal{A}^2}  \sqrt{\bm{k}_\perp^{\prime 2} + \mathcal{A}^2} }.
\end{eqnarray} 
%%%%%%%%%%%%%%%
Note that the EMFFs at the zero momentum transfer are normalized as
%%%%%%%%%%%%%%%%
\begin{eqnarray}
\label{eq:normFF}
    F_M(0) = e_q F^q_M (0)+ e_{\bar{q}} F^{\bar{q}}_M (0),
\end{eqnarray}
%%%%%%%%%%%%%%
where at $Q^2 =0$, the quark sector form factors for the quark $F_M^q (0)$ and antiquark $F_M^{\bar{q}} (0)$ in Eq.~(\ref{eq:normFF}) are defined to be equal to one, respectively. 
%The symbols of $e_q$ and $e_{\bar{q}}$ are the quark and antiquark charges.

The corresponding mean square charge radius of the meson can be computed as
\begin{eqnarray}
    \expval{r_M^2} = -6\left.\frac{\partial F_M(Q^2)}{\partial Q^2}\right|_{Q^2\to 0},
\end{eqnarray}
where we can also decompose it by its quark flavor as
\begin{eqnarray}
   \expval{r_M^2} = e_q \expval{r_{M,q}^2} + e_{\bar{q}} \expval{r_{M,\bar{q}}^2}.
\end{eqnarray}
Therefore, with this expression, we can also study the contribution of each quark flavor separately.

%================================================================
\subsection{EMFFs in nuclear medium}
%================================================================

When mesons are in a nuclear medium, the light quark mass is directly affected by the Lorentz scalar potential, and the light quark energy is directly influenced by the Lorentz vector potential. We assume the magnitude of vector potential is the same for all mesons, which contain one light quark or light antiquark. The increased repulsion of the $K^+$ meson can be explained by a reduction in the strong coupling constant $\alpha_s$ in the medium, as discussed in Ref.~\cite{Arifi:2023jfe}. In the QMC model, adjusting the vector potential is necessary to match the empirically extracted slightly repulsive total potential of the $K^+$ meson, as shown in Ref.~\cite{Tsushima:1997df}.

We now investigate how the in-medium modifications of the quark properties take place, as modeled in the QMC model, and affect the meson properties such as the EMFFs in the LFQM. The two key inputs from the QMC model are:
%%%%%%%%%%%%%%%%
\begin{enumerate}
    \item The in-medium effective mass of light quarks, modified by the scalar $\sigma$ mean-field.
    \item The in-medium energy of light quarks, modified by the vector potential.
\end{enumerate}
%%%%%%%%%%%%%%
Note that the Gaussian parameter $\beta_{q\bar{q}}$ may be modified in the medium; however, for simplicity, we assume it to be constant in the present work, as it is expected to remain and not largely changed in the medium because it is associated with the short-range scale of the meson wave function. 
The effects of the medium modifications on the light and heavy-light mesons depend on their quark constituents, which we will explain further.

The mass of the light quark or antiquark is modified by the scalar potential as
\begin{eqnarray}
m_q^* &=& m_q - V_{q\sigma}.
\end{eqnarray}
The energy of the light quark and antiquark, $p^{*0}_{i}$, is modified by the vector potential
\begin{eqnarray}
p^{*0}_{i} &=& 
\begin{cases}
E_q^* + V_{q\omega} & \text{for light quark,} \\
E_{\bar{q}}^* - V_{q\omega} & \text{for light antiquark.}
\end{cases}
\end{eqnarray}
where $E_q^* = E_{\bar{q}}^* = \sqrt{m_q^{*2} + \bm{p}^2_q}$.
The total energy of a meson, $P^{*0}$, is then given by
\begin{eqnarray}
P^{*0} &=& 
\begin{cases}
E_M^* & \text{for } (q\bar{q}), \\
E_M^* + V_{q\omega} & \text{for } (q\bar{Q}), \\
E_M^* - V_{q\omega} & \text{for } (Q\bar{q}),
\end{cases}
\end{eqnarray}
where $E_M^* = \sqrt{M^{*2} + \bm{P}^2}$, with $M^{*}$ being the in-medium meson mass. The vector potential cancels for $(q\bar{q})$ mesons, and there is no direct effect from the vector potential on $Q$ and $\bar{Q}$. The variable $x$, representing the ratio of quark to meson momenta in free space, is defined as
\begin{eqnarray}
x &=& \frac{p^+_q}{P^+} = \frac{p^0_q + p^3_q}{P^0 + P^3} = \frac{E_q + p^3_q}{E_M + P^3}
\end{eqnarray}
This variable $x$ will be modified in the nuclear medium by the scalar and vector potentials.

%%---------------------------------------------------------------
\subsubsection{Equal quark mass case}
%%---------------------------------------------------------------

For the $q\bar{q}$ mesons, the longitudinal momentum of the quark (antiquark) is modified by the vector potential $+ V_{q\omega} (- V_q{\omega})$, with the new definition of the ``quark" longitudinal momentum given by~\cite{Arifi:2023jfe}
%%%%%%%%%
\begin{eqnarray}\label{eq:long}
x &\to&
 \tilde{x}^*
= \frac{p_q^{*+} +  V_{q\omega} }{P^{*+}} = x^* + \frac{V_{q\omega} }{P^{*+}},
\end{eqnarray}
The vector potentials for the quark and antiquark cancel out in the case of the $q\bar{q}$ mesons. In this new definition, the longitudinal momentum $x$ is shifted due to the difference between $(p_q^+, P^+)$ and $(p_q^{*+}, P^{*+})$ when computing the form factor in the medium. 
The quark sector form factor due to the quark (not the antiquark) contribution is then calculated as
%%%%%%%%%%%%%%%
\begin{eqnarray}
F_{M}^{i *} (Q^2) &=& \int_{-\frac{V_{q\omega} }{P^{*+}}}^{1-\frac{V_{q\omega} 
}{P^{*+}}} \dd x^*  \int \frac{\dd ^2\bm{k}_\perp}{2(2\pi)^3} \nonumber\\ 
&& \times {\Phi}(\tilde{x}^*, \bm{k}_\bot) {\Phi}^\prime(\tilde{x}^*, \bm{k}_\bot^\prime) \nonumber\\
&& \times 
 \frac{  (\bm{k}_\perp\cdot\bm{k}_\perp^\prime + \mathcal{A}(\Tilde{x}^*)^2)}{\sqrt{\bm{k}_\perp^2 + \mathcal{A}(\Tilde{x}^*)^2}  \sqrt{\bm{k}_\perp^{\prime 2} + \mathcal{A}(\Tilde{x}^*)^2} }.\quad \quad 
\end{eqnarray} 
%%%%%%%%%%%%%%
After shifting the integration limits by ${V_{q\omega}}/{P^{*+}}$, the result remains the same because the limits cancel each other out. 
In a similar manner, one can calculate 
the antiquark contribution to the EMFFs.
The final expression for the quark contribution to the EMFFs in the medium, using $\tilde{x}^*$, is
%%%%%%%%%%%%%%%%
\begin{eqnarray}\label{eq:emff_eequal}
     F_{M}^{i*}(Q^2) &=&  \int_{0}^{1} \dd \tilde{x}^* \int \frac{\dd ^2\bm{k}_\perp}{2(2\pi)^3} {\Phi}(\tilde{x}^*, \bm{k}_\bot) {\Phi}^\prime(\tilde{x}^*, \bm{k}_\bot^\prime) \nonumber\\ 
    && \times 
 \frac{ (\bm{k}_\perp\cdot\bm{k}_\perp^\prime + \mathcal{A}^{*2}) }{\sqrt{\bm{k}_\perp^2 + \mathcal{A}^{*2}}  \sqrt{\bm{k}_\perp^{\prime 2} + \mathcal{A}^{*2}} },\quad \quad 
\end{eqnarray} 
%%%%%%%%%%%%%%% 
where $\mathcal{A}^{*} \equiv \mathcal{A}(\tilde{x}^*)$ is taken for simplicity.
%and $\Phi_i^* = \Phi_i(\tilde{x}^*, \bm{k}_\perp)$, $\Phi_f^{*'} = \Phi_f(\tilde{x}^*, \bm{k}_\perp')$. 
This equation shows that the scalar potential affects the EMFFs through the effective quark mass $m_q^*$~\cite{deMelo:2014gea}, while the vector potential modifies the energies, but the effects cancel out between the quark and antiquark.

%%---------------------------------------------------------------
\subsubsection{Unequal quark mass case}
%%---------------------------------------------------------------
For the $q\bar{Q}$ and $Q\bar{q}$ mesons, the scalar and vector potentials affect both the meson four-momenta.
%and the meson weak-decay constants. 
The longitudinal momenta of the quark and antiquark in the medium are given by
%%%%%%%%%%%%%%%%
\begin{eqnarray}
x &\to& 
\begin{cases}
 \tilde{x}^* = \dfrac{p_q^{*+} + V_{q\omega} }{P^{*+} + V_{q\omega}} =  \dfrac{x^* +
V_{q\omega}/P^{*+}}{(1 + 
V_{q\omega}/P^{*+})} & {\rm for\ } (q\Bar{Q}),\\ 
 \\
  \tilde{x}^* = \dfrac{p_q^{*+} - V_{q\omega} }{P^{*+} - V_{q\omega}} =
\dfrac{x^*  - V_{q\omega}/P^{*+}
}{(1 - V_{q\omega}/P^{*+}) }  
  & {\rm for\ } (Q\Bar{q}). \nonumber
\end{cases}\\
\end{eqnarray}
%%%%%%%%%%%%%%
In such cases, we can define the variable of the integration by
%%%%%%%%
\begin{eqnarray}
\dd x^* = (1 \pm V_{q\omega}/P^{*+})\ \dd \tilde{x}^*.
\end{eqnarray}
Thus, using this transformed variable, the EMFFs for the $q\bar{Q}$ and $Q\bar{q}$ mesons in the medium can be written as
%%%%%%%%
\begin{eqnarray}\label{eq:vector}
 F_{M}^{i*}(Q^2) &=&  \int_{0}^{1} \dd \tilde{x}^* \int \frac{\dd ^2\bm{k}_\perp}{2(2\pi)^3} \left(1 \pm \frac{V_{q\omega}}{P^{*+}}\right) \nonumber\\ 
    & & \times  {\Phi}(\tilde{x}^*, \bm{k}_\bot) {\Phi}^\prime(\tilde{x}^*, \bm{k}_\bot^\prime)\nonumber\\
 && \times  \frac{ (\bm{k}_\perp\cdot\bm{k}_\perp^\prime + \mathcal{A}^{*2}) }{\sqrt{\bm{k}_\perp^2 + \mathcal{A}^{*2}} \sqrt{\bm{k}_\perp^{\prime 2} + \mathcal{A}^{*2}} }.
\end{eqnarray} 
The final expressions for the EMFFs of the heavy-light mesons in the medium, considering the vector potential, are similar to those for the quark distributions in the nuclear medium with a vector potential as derived in Ref.~\cite{Steffens:2004yb}. 
In the meson rest frame, the total momentum is $P^{*+} = \sqrt{M^{*2} + \bm{P}^2} + P^{*3} = M^*$. 
In our LFQM, the meson mass is replaced by the invariant mass $M_0^*$, to be consistent with the BT construction.
However, if we take the absolute values and their average values of the meson and antimeson, the contribution of the vector potential will be canceled out, leaving only the modifications from the scalar potential~\cite{Arifi:2023jfe}.  (For example, taking the average (of absolute) values of the $K^+$ and $K^-$ EMFFs.)

%%%%%%%%%%%%%%%
\begin{table}[b]
	\begin{ruledtabular}
		\renewcommand{\arraystretch}{1.3}
		\caption{The quark masses $m_{q(Q)}$ and scale parameters $\beta$ used in the present work, obtained with linear confining potential adopted from Ref.~\cite{Choi:1997iq}. (All values are in GeV.)  }
		\label{tab:parameter}
		\begin{tabular}{cccccccc}
		  $m_q$ & $m_s$ & $m_c$ & $m_b$  &	$\beta_{q\bar{q}}$ & $\beta_{q\bar{s}}$ & 
		 $\beta_{q\bar{c}}$ & $\beta_{q\bar{b}}$  \\ \hline
	    0.22 & 0.45 & 1.80 & 5.20 & 0.3659 & 0.3886 & 0.4679 & 0.5266 \\
          \end{tabular}
		\renewcommand{\arraystretch}{1}
	\end{ruledtabular}
\end{table}
%%%%%%%%%%%%%%%%

%%%%%%%%%%%%%%%%%
\begin{table}[t]
	\begin{ruledtabular}
		\renewcommand{\arraystretch}{1.3}
		\caption{Bag parameters used in this study, fitted to match the nucleon mass and radius in free space. }
		\label{tab:bag}
		\begin{tabular}{ccccc}
		$m_q$ [MeV]  &	$B^{1/4}$ [MeV] & $Z_N$ & $x_q$ & $S_N(\sigma=0)$\\ \hline
		220 & 148 & 4.327 & 2.368 & 0.609 \\
		\end{tabular}
		\renewcommand{\arraystretch}{1}
	\end{ruledtabular}
\end{table}
%%%%%%%%%%%%%%

%%%%%%%%%%%%%%%%%%%%%%%%%%
\begin{table}[t]
	\begin{ruledtabular}
    		\renewcommand{\arraystretch}{1.3}
		\caption{Coupling constants and the incompressibility $K$ obtained in the QMC model.  ($m_N^*$ value is at $\rho_0 = 0.15$ fm$^{-3}$.)}
		\label{tab:coupling}
		\begin{tabular}{ccccc}
		$m_q$ [MeV] &	$g_{N\sigma}^2/4\pi$ & $g_{N\omega}^2/4\pi$ & $m_N^*$ [MeV] &  $K$ [MeV]\\ \hline
		220 & 6.40 & 7.57 & 699 & 321 \\
		\end{tabular}
        		\renewcommand{\arraystretch}{1}
	\end{ruledtabular}
\end{table}
%%%%%%%%%%%%%%%%%%%%%

%%%%%%%%%%%%%%%%%%
\begin{figure}[b]
	\centering
	\includegraphics[width=0.9\columnwidth]{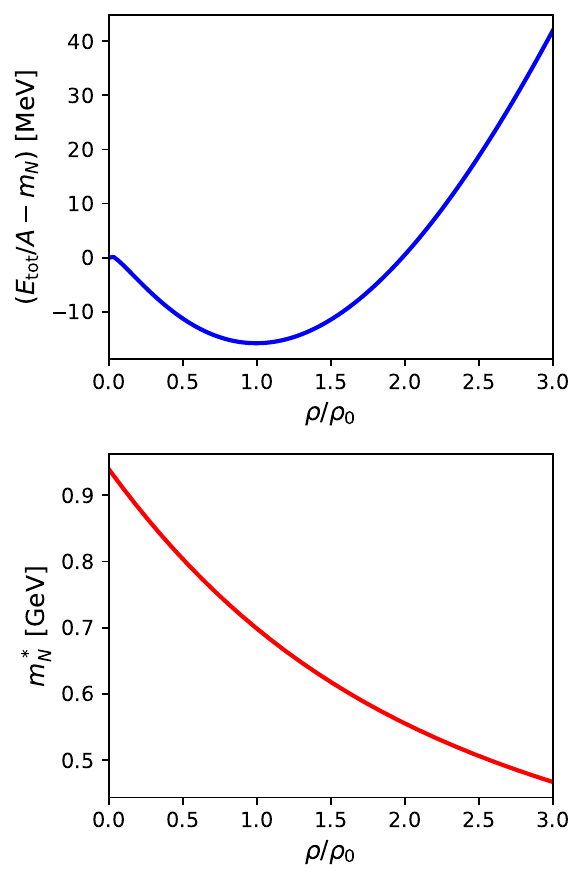}
 	\caption{\label{fig:eos} Density dependence of total energy per nucleon minus $m_N$, $E_{\rm tot}/A - m_N$ (upper panel), and effective nucleon mass $m_N^*$ (lower panel). } 
\end{figure}
%%%%%%%%%%%%%%

%================================================================
\section{Numerical results} \label{sec:num}
%================================================================

This section presents our numerical results for the in-medium EMFFs of the light and heavy-light pseudoscalar mesons in SNM. 
Before discussing the results, we first explain the model parameters used in the LFQM and QMC models.

%================================================================
\subsection{Model parameters}
%================================================================

The LFQM parameters used in this study are given in Table~\ref{tab:parameter}. 
They show good agreement with the meson masses and decay constants in free space, as calculated by Choi and Ji~\cite{Choi:1997iq}. 
With these parameters, we explore the EMFFs of the light and heavy-light mesons in SNM, using the QMC model inputs for the in-medium light quark properties. The same parameters were also used in our previous work on the in-medium meson decay constants and DAs~\cite{Arifi:2023jfe}.

%%%%%%%%%%%%%%%%
\begin{figure}[t]
	\centering
	\includegraphics[width=0.9\columnwidth]{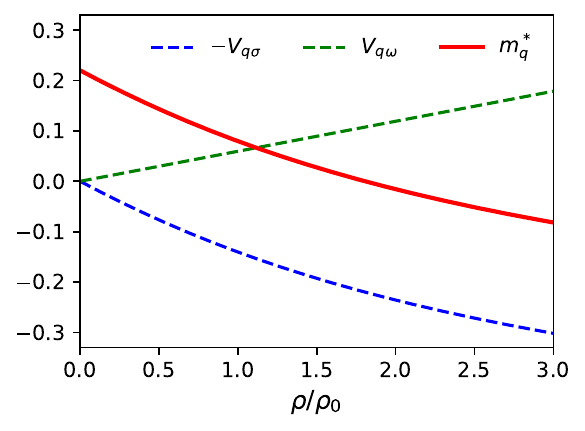}
 	\caption{\label{fig:pot} Density dependence of the scalar (blue dashed line) and vector (green dashed line) mean-field potentials ($-V_{q\sigma}, V_{q\omega})$ and the light-quark effective mass $m_q^*$ (red solid line). } 
\end{figure}
%%%%%%%%%%%%%

%%%%%%%%%%%%%%%%%
\begin{figure}[b]
	\centering
	\includegraphics[width=0.9\columnwidth]{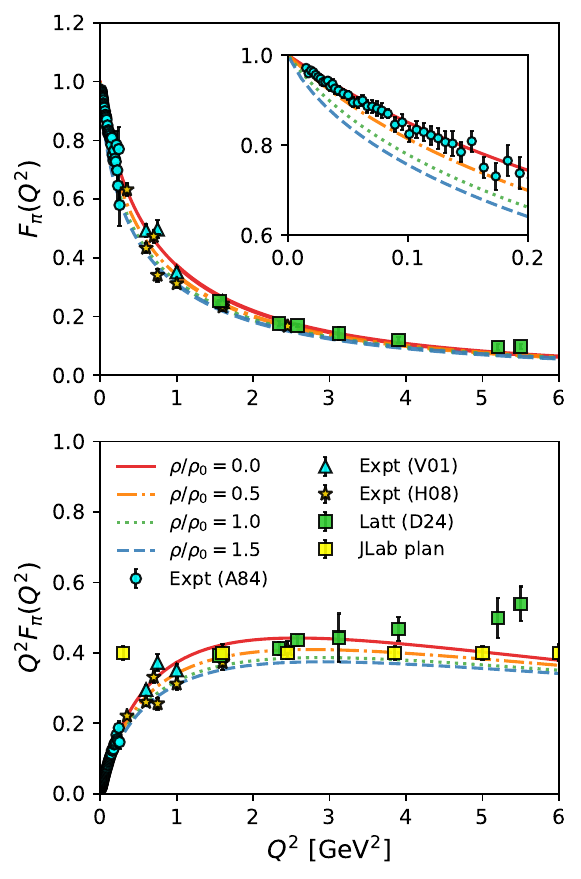}
 	\caption{\label{fig:ffqq} In-medium EMFFs of the pion (upper panel) and $Q^2 F_{\pi} (Q^2)$ (lower panel) for a few nuclear densities compared with available experimental data~\cite{Amendolia:1984nz, JeffersonLabFpi:2000nlc,JeffersonLab:2008jve} and recent lattice QCD data~\cite{Ding:2024lfj} of the pion EMFFs in free space. 
    The expected coverage of the forthcoming JLab measurement~\cite{Horn:2019} are also shown, where the central points were chosen arbitrarily. 
    The fall-off of the pion EMFFs near $Q^2=0$ becomes more pronounced at higher density, resulting a considerable change of the pion's charge radius in nuclear medium.} 
\end{figure}
%%%%%%%%%%%%%%%%

%%%%%%%%%%%%%%%%%
\begin{figure}[t]
	\centering
	\includegraphics[width=0.9\columnwidth]{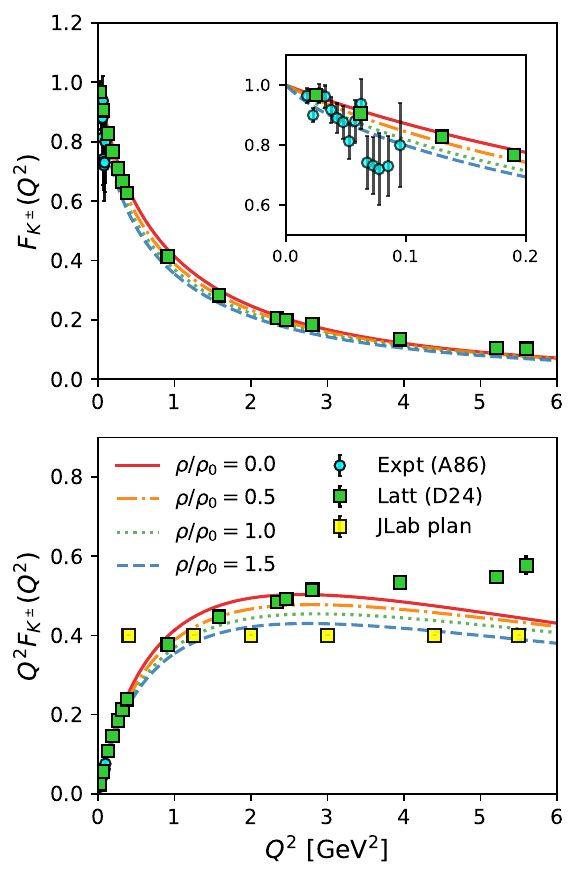}
 	\caption{\label{fig:ffqs_charged} In-medium EMFFs of the charged kaon (upper panel) and $Q^2 F_{K^\pm} (Q^2)$ (lower panel) with a few densities 
    compared with available experimental data~\cite{Amendolia:1986ui} and recent lattice QCD data~\cite{Ding:2024lfj} of EMFFs in the free space. The expected coverage of the forthcoming JLab measurement~\cite{Horn:2009} are also shown, where the central points were chosen arbitrarily.}
\end{figure}
%%%%%%%%%%%%%%%%

%%%%%%%%%%%%%%%%%
\begin{figure}[t]
	\centering
	\includegraphics[width=0.9\columnwidth]{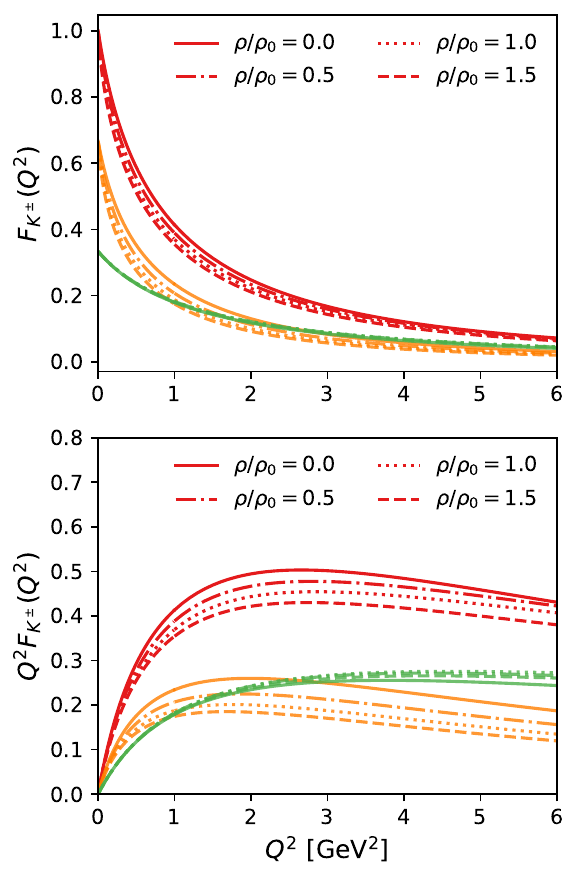}
 	\caption{\label{fig:ffqs_charged_quark} In-medium total EMFFs (red), along with the up quark (orange) and strange quark (green) form factors of the charged kaon multiplying by the corresponding quark charges $e_q$ as a function of $Q^2$ for a few densities [upper panel]. $Q^2 F_{K^\pm}$ along with its quark form factors are shown with various densities [lower panel].} 
\end{figure}
%%%%%%%%%%%%%%%

We now describe the standard QMC model parameters based on the MIT bag model~\cite{Saito:2005rv}. 
The quark-meson coupling constants are determined by fitting the empirical saturation properties of SNM. 
In the QMC model, the bag parameters ($Z_N$ and $B$) are set by the free nucleon mass $m_N = 939$ MeV and the bag radius $R_N = 0.8$ fm, 
along with the mass stability condition. 
These parameters are shown in Table~\ref{tab:bag}, using the light quark mass of $m_q = 220$ MeV.

Next, we present the negative of the binding energy per nucleon, $E_{\rm tot}/A - m_N$, for $m_q = 220$ MeV. 
The nucleon-scalar and nucleon-vector coupling constants, $g_\sigma^N$ and $g_\omega^N$, are determined by fitting the negative of the binding energy $-15.7$ MeV at saturation density $\rho_0 = 0.15$ fm$^{-3}$ ($k_F = 1.305$ fm$^{-1}$), as shown in Table~\ref{tab:coupling}. 
The density dependence of $E_{\rm tot}/A - m_N$ is shown in the upper panel of Fig.~\ref{fig:eos}.

Figure~\ref{fig:eos} shows the total energy per nucleon minus $m_N$, $E_{\rm tot}/A - m_N$ (upper panel), and the effective nucleon mass (lower panel) as a function of the nuclear density ratio $\rho/\rho_0$ for a light quark mass in free space of $m_q = 220$ MeV. The calculated value of the incompressibility coefficient $K = 321$ MeV is near the empirical range of $K = 200-300$ MeV~\cite{Stone:2014wza}, though slightly higher.

The mean-field potentials for the light quarks, which act on the light quarks in mesons, are shown in Fig.~\ref{fig:pot}. Using these mean-field potentials and effective (anti)quark masses, we study the light and heavy-light meson properties in SNM. The in-medium decay constants and DAs are discussed in our previous work~\cite{Arifi:2023jfe}. 
In this study, we focus on the in-medium modifications of the light and heavy-light meson EMFFs.

%----------------------------------------------------------------
\subsection{In-medium pion EMFFs}
%----------------------------------------------------------------

In Fig.~\ref{fig:ffqq}, we present the results of the pion EMFF for a few different nuclear densities, 
comparing with experimental data~\cite{Amendolia:1984nz, JeffersonLabFpi:2000nlc,JeffersonLab:2008jve} and recent lattice QCD data~\cite{Ding:2024lfj} in free space. 
The upper and lower panels in Fig.~\ref{fig:ffqq} respectively illustrate 
$F_\pi(Q^2)$ and $Q^2 F_\pi(Q^2)$ plotted up to $Q^2=6$ GeV$^2$, which is the highest accessible $Q^2$ point at JLab. 
It is worth noting that we display only the positively charged pion ($\pi^+$) results. 
In free space, the results for the pion EMFF in the low-$Q^2$ region exhibit 
good agreement with the existing experimental data~\cite{Amendolia:1984nz, JeffersonLabFpi:2000nlc,JeffersonLab:2008jve}, which indicates the reliability of the model to apply to the nuclear medium. 
However, at $Q^2$ around 4–6 GeV$^2$, there is some discrepancy with lattice QCD~\cite{Ding:2024lfj}, although this should be tested in the upcoming JLab experiment.

%%%%%%%%%%%%%%%%%%
\begin{figure}[t]
	\centering
	\includegraphics[width=0.9\columnwidth]{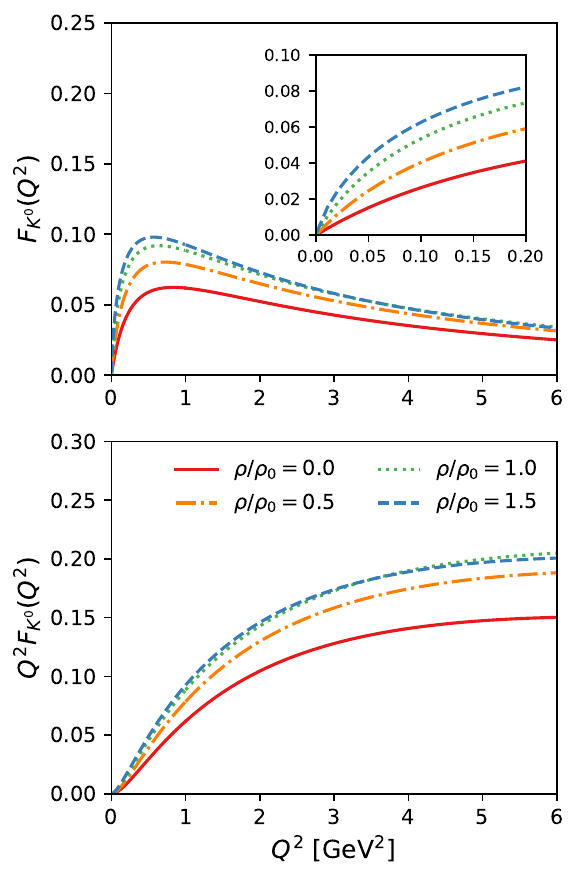}
 	\caption{\label{fig:ffqs_neutral} In-medium EMFFs of the neutral kaon (upper panel) and $Q^2 F_{K^0} (Q^2)$ (lower panel) for a few densities. } 
\end{figure}
%%%%%%%%%%%%%%%%%%

%%%%%%%%%%%%%%%%%%
\begin{figure}[t]
	\centering
	\includegraphics[width=0.9\columnwidth]{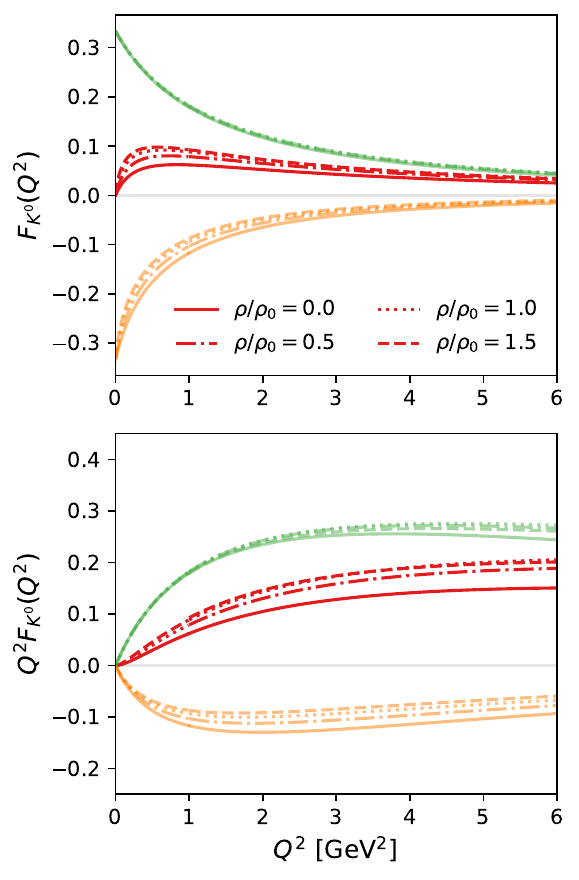}
 	\caption{\label{fig:ffqs_neutral_quark} In-medium total EMFFs (red), along with down quark (orange) and strange quark (green) form factors of the neutral kaon multiplying by the corresponding quark charges $e_q$ as a function of $Q^2$ for a few densities [upper panel]. $Q^2 F_{K^0}$ along with its quark form factors are shown with a few densities [lower panel].} 
\end{figure}
%%%%%%%%%%%%%%%%

When examining the pion EMFF in symmetric nuclear matter (SNM), only the scalar potential influences the in-medium modifications, as the light quark and antiquark vector potentials cancel each other out. In Eq.~\eqref{eq:emff_eequal}, the operator is given by two terms
\begin{eqnarray}
 \mathcal{O}_{EM} \propto (\bm{k}_\perp\cdot\bm{k}_\perp^\prime + m_q^{*2}),
\end{eqnarray}
where the value of $m_q^{*2}$ is reduced in the nuclear medium due to the scalar potential. However, the first term, corresponding to the transverse motion of the quark, is not modified in the medium. 
Therefore, even though $m_q^*$ is significantly modified in the medium, the nuclear modifications to the EMFF are constrained, but remain significant in the low-$Q^2$ region.
The upper panel of Fig.~\ref{fig:ffqq} clearly shows that the difference $|F_\pi^* - F_\pi|$ reaches its maximum in the low $Q^2$ region, around 0.5–1.0 GeV$^2$, while at higher $Q^2$, the differences become less pronounced. 
The results are reflected in the lower panel of Fig.~\ref{fig:ffqq}, 
where $Q^2 F_\pi(Q^2)$ converges at large $Q^2$, indicating that the impact of medium effects diminishes at higher momentum transfers (asymptotic QCD region).

As nuclear density increases, the pion EMFF exhibits a more pronounced fall-off, particularly in the low-$Q^2$ region, as illustrated in Fig.~\ref{fig:ffqq}.  
At $\rho/\rho_0=1$ or 1.5, the in-medium EMFF decreases more rapidly than the experimental data near $Q^2=0$, as shown in the inset of Fig.~\ref{fig:ffqq}.  
In this region, the experimental measurements are quite precise, allowing clear distinction of medium modifications to the EMFFs.  
Consequently, the pion’s charge radius undergoes significant changes, increasing to approximately 1.45 times its value in free space, with $\expval{r^{*2}_\pi} = 0.897 \,\text{fm}^2$ at $\rho/\rho_0 = 1$, which is significantly larger than the experimental value of $0.427(10) \,\text{fm}^2$ in free space~\cite{Amendolia:1986ui}. 
The density dependence of this modification will be further discussed in Sec.~\ref{sec:rad}.

At higher $Q^2$, however, in-medium modifications become smaller than the experimental uncertainties.  
Moreover, the discrepancies between different experimental and lattice data remain substantial, making it challenging to detect medium effects at this stage.  
%Additionally, uncertainties in the parameters of the LFQM can influence the EMFF, further complicating the analysis. 
More precise experimental data, such as those anticipated from future measurements at JLab, will be crucial for providing an accurate description of the EMFF in free space and for placing stronger constraints on the model.

It is worth noting that there are also several model uncertainties that affect the EMFF predictions.
From the QMC model, one key uncertainty arises from the quark mass dependence in the nuclear medium.
If the quark mass decreases more rapidly in medium, the EMFF will also decrease more quickly, whereas a slower mass reduction leads to a more gradual decrease.
Another source of uncertainty comes from the LFQM parameters, which influence the shape of the EMFF in free space.
However, the main parameters, $m_q$ and $\beta$, are typically constrained to reproduce observables such as the mass, decay constants, and charge radius. 
As a result, model uncertainties are smaller in the low-$Q^2$ region but grow with increasing $Q^2$~\cite{Choi:1997iq}.

Despite the difficulty in making precise quantitative predictions for the EMFF, the trend of a more rapid fall-off in the nuclear medium remains a robust feature.  
Furthermore, examining medium effects through the charge radius or the EMFF behavior near $Q^2=0$ provides a clearer indication of in-medium modifications.
This behavior is consistent with the expected effects of increased interactions within the nuclear medium, which lead to an expanded spatial charge distribution of the pion.

In this study, we have limited our study of the in-medium EMFFs to densities up to $\rho = 1.5\rho_0$, 
since at higher densities, the pion decay constant may become negative, as already discussed in Ref.~\cite{Arifi:2023jfe}. Thus, we regard the limitation of the present approach for the higher nuclear matter densities. The results of the pion EMFFs at higher densities up to $3\rho_0$ in the LFQM associated with the QMC model with the help of the BSA, we refer the interested readers to Ref.~\cite{Yabusaki:2023zin}.

A noteworthy comparison can be drawn with the results obtained using the LFQM associated with the QMC model with the help of the BSA. Our findings indicate that the fall-off of the EMFFs at $\rho = \rho_0$ is slower compared to the predictions from the LFQM combined with the QMC model with the help of the BSA~\cite{deMelo:2014gea}. This difference highlights the sensitivity of the pion's EMFFs to the treatment of the medium effects, and it suggests that the present approach captures different aspects of the in-medium modifications.

%%%%%%%%%%%%%%%%%
\begin{figure*}[t]
	\centering
	\includegraphics[width=1.8\columnwidth]{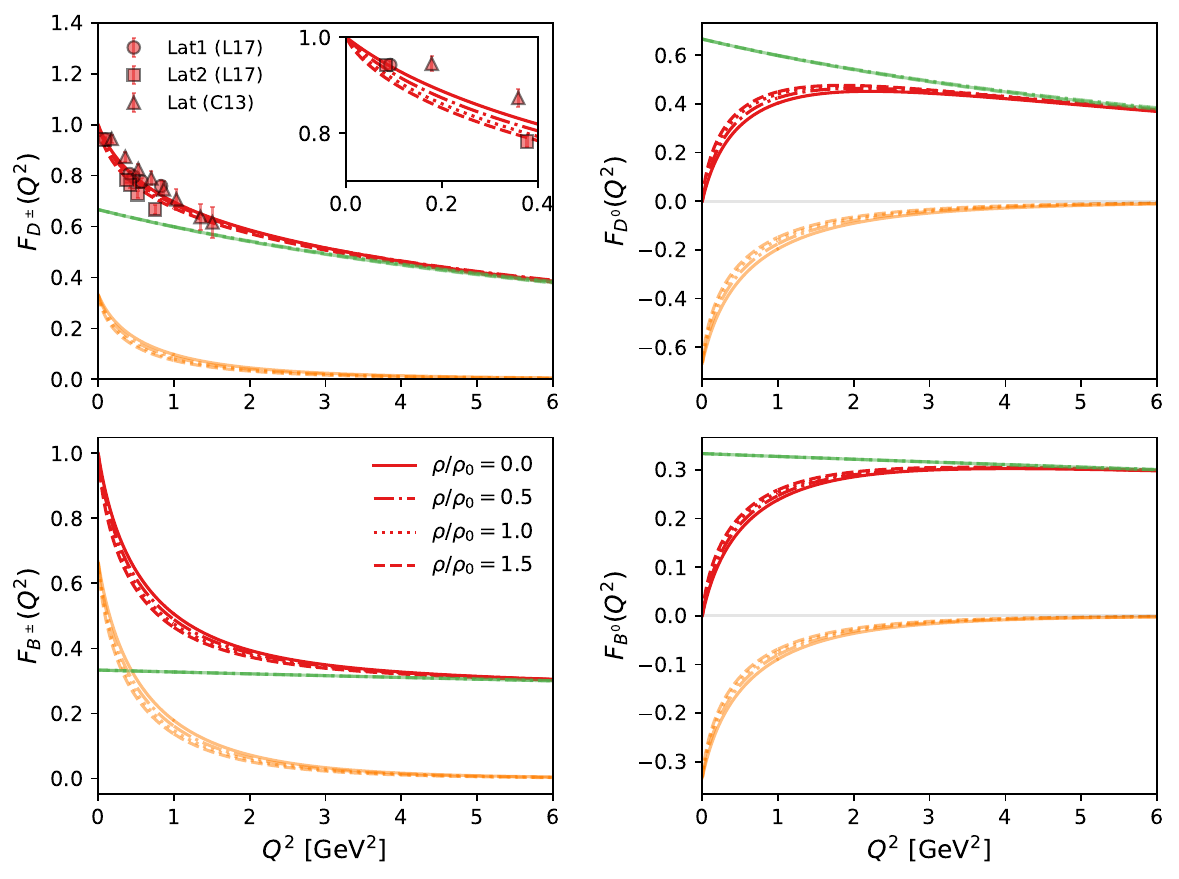}
 	\caption{\label{fig:ffqQ} In-medium EMFFs of the charged and neutral $D$ and $B$ mesons with a few densities along with its quark sector form factors of the light quark (orange) and heavy quark (green). In free space, the $D$ meson EMFF is consistent with the lattice QCD simulations~\cite{Can:2012tx,Li:2017eic}. In the nuclear medium (SNM), we show that the heavy quark contribution $e_Q F^Q_{M} (Q^2)$ is almost unchanged. The main nuclear modifications originate from the light quark $e_q F^q_{M} (Q^2)$. } 
\end{figure*}
%%%%%%%%%%%%%%%

%----------------------------------------------------------------
\subsection{In-medium kaon EMFFs}
%----------------------------------------------------------------

For the EMFFs of kaon, in general, we have nonvanishing $Q^2$ dependence of the charged and neutral kaon form factors.
Due to different quark contents, up/down, and strange quarks, the vector potential gives different effects for $K^+(u\bar{s})$ and $K^-(s\bar{u})$ as observed in our previous study of in-medium decay constants~\cite{Arifi:2023jfe}.
However, here we take the absolute values of the form factors and their average values which will cancel the contribution of the vector potential, leaving only the medium modifications from the scalar potential to play the role in the form factors. The reason for this treatment is that one can get a unified picture of the in-medium modifications focusing on the internal structure change or the important dynamics, and can avoid the vector-potential based "shift" of EMFFs, such as the energy (four-momentum) shift.
However, we will discuss the effect of the vector potential on the charge radius of kaons and other mesons with unequal quark contents in more detail in Sec.~\ref{sec:rad}, as it is the most significantly impacted observable in the nuclear medium.

%----------------------------------------------------------------
\subsubsection{Charged kaon}
%----------------------------------------------------------------

Figure~\ref{fig:ffqs_charged} presents the results for the $K^{\pm}$ EMFFs alongside the older experimental data~\cite{Amendolia:1986ui} and recent lattice QCD results~\cite{Ding:2024lfj}. 
As shown in the inset of the upper panel, the model agrees well with the lattice data in the low-$Q^2$ region, though some discrepancies persist at higher $Q^2$.
We also found that the in-medium kaon EMFFs decrease with increasing nuclear density, similar to the pion's case. 
For a clear illustration of the EMFF behaviors in the nuclear medium, we also show the results for $Q^2 F_{K^{\pm}} (Q^2)$ in the lower panel of Fig.~\ref{fig:ffqs_charged}.

In Fig.~\ref{fig:ffqs_charged_quark}, we also show the total kaon EMFFs and their quark sector form factors as a function of $Q^2$ for a few different nuclear densities. 
In the upper panel of Fig.~\ref{fig:ffqs_charged_quark}, we observe that the up quark form factor $e_u F^u_{M}(Q^2)$, (blue line) decreases with increasing nuclear density and $Q^2$, while the strange quark form factor $e_s F^s_{M}(Q^2)$ (red line) slightly increases with density.
This is because the strange quark does not directly interact with the nuclear medium (meson mean fields) in the kaon.
As complementary, we also show our results for the $Q^2 F_{K^{\pm}} (Q^2)$ as a function of the nuclear density in the lower panel of Fig.~\ref{fig:ffqs_charged_quark}, where the nuclear medium effects can be seen more clearly for the total kaon EMFFs and their quark sector form factors.

%----------------------------------------------------------------
\subsubsection{Neutral Kaon}
%----------------------------------------------------------------

Besides computing the $K^{\pm}$ EMFFs, we also compute the $K^0$ EMFFs in the nuclear medium as a function of the $Q^2$ for a few different nuclear densities as shown in Fig.~\ref{fig:ffqs_neutral}. 
The upper panel of Fig.~\ref{fig:ffqs_neutral} shows that the total $K^0$ EMFFs increase as the density increases, which is different from what we found in the $K^{\pm}$ EMFFs. 
%This indicates that the scalar potential impacts positively on the $K^0$ EMFFs. 
A similar indication is found in the lower panel of Fig.~\ref{fig:ffqs_neutral}. 
It shows that the increase is more pronounced at the higher nuclear density. 
Also, our results on the $K^0$ EMFFs are consistent with other model predictions obtained as in Ref.~\cite{Gifari:2024ssz}.

Moreover, we also observe the contributions of each quark sector form factor to the total $K^0$ EMFF as illustrated in Fig.~\ref{fig:ffqs_neutral_quark}. 
In the upper panel of Fig.~\ref{fig:ffqs_neutral_quark}, we show that the form factors of both the down quark $e_d F^d_{M}(Q^2)$  (blue line) and strange quark $e_s F^s_{M}(Q^2)$ (red line) increase as the density increases. 
However, it is evident that the light quark sector form factor contributes significantly to the $K^0$ EMFF, particularly in the lower $Q^2$ region.
The clear contributions of each quark form factor for the $K^0$ is shown in the lower panel of Fig.~\ref{fig:ffqs_neutral_quark}.

%----------------------------------------------------------------
\subsection{In-medium $D$ and $B$ mesons EMFFs}
%----------------------------------------------------------------

Figure~\ref{fig:ffqQ} presents the EMFFs of the $D$ and $B$ pseudoscalar mesons, their average values of the absolute form factors along with their quark sector form factors. 
In free space, the model predictions for the $D$ meson are consistent with lattice QCD data~\cite{Can:2012tx,Li:2017eic}, although the uncertainties among the data remain large.
In a nuclear medium, we find that the contributions from the heavy quark sector form factors, $e_Q F^Q_{M}(Q^2)$, remain almost unchanged, while the primary nuclear modifications arise from the light quark sector form factors, $e_q F^q_{M}(Q^2)$. The figure also shows that the fall-off of the in-medium $B^+$ meson EMFFs is faster than that of the $D^+$, as the light quark appears as an antiquark in the $D^+$, suppressing its contribution. Conversely, the increase in the in-medium $D^0$ meson EMFFs is faster than that of the $B^0$ mesons for a similar reason.
Furthermore, in the high-$Q^2$ region, the form factor is dominated by the heavy quark contribution $e_Q F^Q_{M}(Q^2)$, resulting in negligible medium modifications.

%%---------------------------------------------------------------
\subsection{In-medium charge radius}
\label{sec:rad}
%%---------------------------------------------------------------

The charge radii of the charged and neutral pseudoscalar mesons are presented in Table~\ref{tab:radius}. For the positively charged mesons, the radius has a positive sign, while for the negatively charged and neutral mesons, it carries a negative sign. 
In this analysis, we focus on the absolute values of the charge radii to discuss their general feature.
In Table~\ref{tab:radius}, as a first step, we consider only the averaged radii of mesons with different charges (e.g., $K^+$ and $K^-$), which are affected by the presence of the vector potential in the nuclear medium.

%%%%%%%%%%%%%%%%%%%%%%%%%%
\begin{table}[b]
	\begin{ruledtabular}
    		\renewcommand{\arraystretch}{1.1}
		\caption{Absolute value of charge radius $|\expval{r^2}|$ of charged and neutral mesons in a nuclear medium in a unit of fm$^2$. The numerical values represent the averaged radius, modified only by the scalar potential. } 
		\label{tab:radius}
		\begin{tabular}{ccccc}
		$\rho/\rho_0$ & $\pi^\pm$ & $K^\pm$ & $D^\pm$ & $B^\pm$ \\ \hline 
            0.00    & 0.428 & 0.360 & 0.184 & 0.354 \\
            0.25    & 0.505 & 0.415 & 0.211 & 0.407 \\
            0.50    & 0.602 & 0.485 & 0.245 & 0.474 \\
            0.75    & 0.728 & 0.576 & 0.288 & 0.557 \\
            1.00    & 0.897 & 0.697 & 0.346 & 0.666 \\
            1.25    & 1.136 & 0.871 & 0.427 & 0.817 \\
            1.50    & 1.523 & 1.154 & 0.555 & 1.053 \\
            Exp.~\cite{Amendolia:1986ui,Amendolia:1984nz}   & 0.427(10)   & 0.34(5)  & \dots & \dots \\ 
            Lat.~\cite{Can:2012tx,Aoki:2015pba}     &  0.458(15) & 0.380(12) & 0.152(26) & \dots \\ \hline
		  $\rho/\rho_0$ & $K^0$ & $D^0$ & $B^0$ &\\ \hline 
            0.00    & 0.084 & 0.315 & 0.175  & \\
            0.25    & 0.113 & 0.370 & 0.202  & \\
            0.50    & 0.149 & 0.439 & 0.235  & \\
            0.75    & 0.194 & 0.526 & 0.277  & \\
            1.00    & 0.255 & 0.641 & 0.331  & \\
            1.25    & 0.341 & 0.802 & 0.406  & \\
            1.50    & 0.481 & 1.058 & 0.525  & \\
            Lat.~\cite{Aoki:2015pba}        & 0.055(10) & \dots & \dots &  \\
		\end{tabular}
            \renewcommand{\arraystretch}{1}
	\end{ruledtabular}
\end{table}
%%%%%%%%%%%%%%%%%%%%%

The meson charge radii in free space show their consistency with experimental data~\cite{Amendolia:1986ui,Amendolia:1984nz} and lattice QCD data~\cite{Can:2012tx} as shown in Table~\ref{tab:radius}. 
We find that the absolute values of the in-medium charge radius increase with increasing nuclear density, which aligns with previous studies that employed different theoretical approaches~\cite{Yabusaki:2023zin,Hutauruk:2018qku,Hutauruk:2019was,deMelo:2014gea}.
However, it should be kept in mind that there exists a recent theoretical prediction showing the pion charge radius might start to decrease at very high nuclear densities, as indicated in Ref.~\cite{Yabusaki:2023zin}, which is rather different from what is found in a very recent study of Ref.~\cite{Gifari:2024ssz}, although further possible experimental verification may be required to confirm this behavior.

We emphasize that the medium modification of charge radius is substantial, as the EMFFs undergo significant changes in the nuclear medium near $Q^2=0$, as described in Fig.~\ref{fig:ffqq}. 
At normal nuclear density, the pion's charge radius ($\expval{r^{*2}_\pi} = 0.897$ fm$^2$) is significantly larger than the experimental value in free space (0.427(10) fm$^2$)~\cite{Amendolia:1986ui}. A similar trend is observed for the kaon, where its in-medium charge radius ($\expval{r^{*2}_K} = 0.697$ fm$^2$) considerably exceeds the free-space value (0.34(5) fm$^2$).
We also find that the $K^0$ radius in normal nuclear matter ($|\expval{r^{*2}_K}| = 0.255$ fm$^2$) is significantly larger than its free-space value (0.055(10) fm$^2$)~\cite{Aoki:2015pba}.

\begin{figure}[t]
	\centering
	\includegraphics[width=0.9\columnwidth]{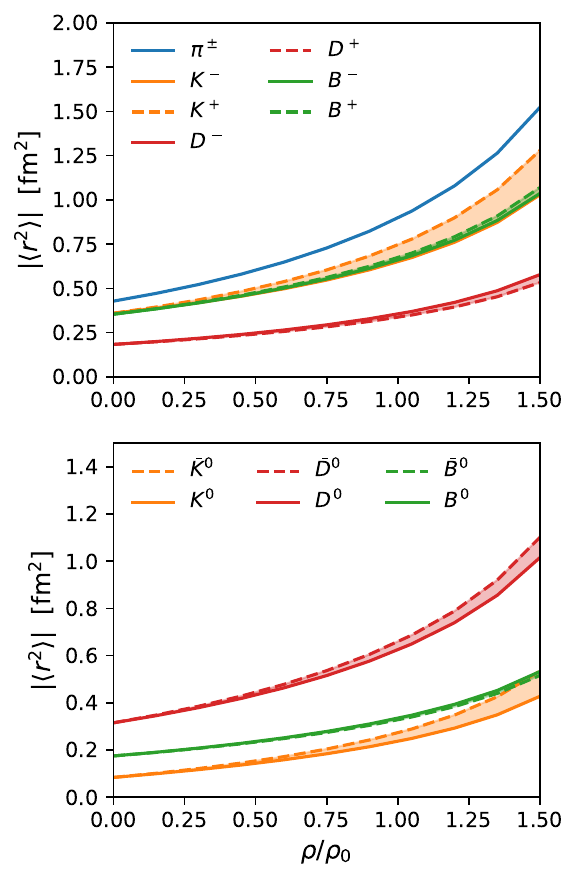} 
  \caption{\label{fig:radius} Density dependence of the charge radii for mesons with different quark flavor contents. The upper panel shows the charged mesons, while the lower panel for the neutral mesons. The absolute value of the charge radius increases with increasing the nuclear density.} 
\end{figure}

Figure~\ref{fig:radius} illustrates that the in-medium charge radius increases differently for various mesons, leading to a clear distinction between the radii of isospin doublets.  
The vector potential plays a crucial role in this effect, causing an increasing difference between, for instance, $\expval{r_{K^+}^2}$ and $\expval{r_{K^-}^2}$ as nuclear density rises.  
For the pion, however, the vector potential cancels out, ensuring that $\pi^+$ and $\pi^-$ maintain identical radii in the medium.  
As expected, the influence of the vector potential is more pronounced for kaons, while its impact on $D$ and $B$ mesons is relatively smaller as also discussed in our previous work~\cite{Arifi:2023jfe}.

Among the charged mesons, the pion's charge radius expands the fastest, whereas the $D$ meson exhibits the slowest growth.  
For neutral mesons, the neutral pion’s charge radius remains zero due to the exact cancellation between its quark and antiquark contributions, which share the same flavor composition ($u\bar{u}$ and $d\bar{d}$).  
In contrast, the neutral $D$ meson displays the most significant charge radius increase among neutral mesons.

Interestingly, despite their different quark contents, the charged kaon and $B$ meson exhibit nearly identical increases in charge radius, with the difference in the vector potential effect arising from its scaling by the invariant mass in $V_{q\omega}/M_0^*$ from Eq.~\eqref{eq:vector}.  
This similarity can be attributed to their quark content, as the $B^+(u\bar{b})$ and $K^+(u\bar{s})$ mesons behave similarly, whereas the $D^+(c\bar{d})$ meson shows a smaller increase, likely due to the presence of a light antiquark instead of a light quark. 
Additionally, the participating quark charges play a role in this behavior, with the absolute charges of the $u$ and $d$ quarks being $2/3$ and $1/3$, respectively.

%%%%%%%%%%%%%%%%%%
%%%%%%%%%%%%%%%%%

%================================================================
\section{Summary} \label{sec:sum}
%================================================================

To summarize, in this study we explored the spacelike electromagnetic form factors (EMFFs) of the light and heavy-light pseudoscalar mesons immersed in the symmetric nuclear matter (SNM). 
To achieve this, we developed a hybrid model integrating a light-front quark model (LFQM) that consistently characterizes meson structure in free space and a nuclear medium. 
Here, we applied the quark-meson coupling (QMC) model to account for the nuclear medium's influence on (light) quarks. 
The properties of the quarks in the medium are modified by self-consistent Lorentz scalar and Lorentz vector mean fields induced by the surrounding nucleons. 

We find that the fall-off (increase) of the in-medium EMFFs of charged (neutral) mesons is faster compared to that in free space. 
The medium modifications of EMFFs are more pronounced in the low-$Q^2$ region.
Consequently, the absolute value of the charge radius of the mesons increases with increasing nuclear density. 
The rate of this increase depends on the meson's quark contents, 
where the pion charge radius growing is the fastest, while that of the charged $D$ meson is the slowest. The charge radii of the kaon and $B$ mesons grow at nearly the same rate. In contrast, for the neutral mesons, the behavior is reversed, where the neutral $D$ meson radius increases more rapidly.

The increase of the meson charge radius in SNM can be attributed to the nuclear modifications in the meson's internal structure and its interactions with the surrounding nuclear environment. 
As nuclear matter density increases, the mesons interact more strongly with the nuclear medium, 
leading to a reduced effective quark mass of the meson's quark-antiquark pair. 
This reduction causes a redistribution of the quark wave functions, increasing the charge radius or the quarks inside mesons immersed in a (higher) density nuclear 
medium influenced by (extra) stronger interactions than those in free space.

Further refinement of the models, such as using more realistic light-front wave functions (LFWFs) and more sophisticated quark models of nuclear matter, and the development of a self-consistent model for calculations in both free space and the nuclear medium, will be important for future studies. Additionally, extending the calculation to analyze other transition form factors and partonic observables such as parton distribution functions (PDFs), generalized parton distributions (GPDs), and transverse momentum distributions (TMDs) of the heavy-light mesons, will provide deeper insight into the in-medium modifications of the meson structure. 
Additionally, establishing connections between the theoretical results with experimental data on the medium modifications will be an important direction for future work.

%================================================================
\section*{Acknowledgement}
%================================================================

The authors would like to express their gratitude for the generous support received during their visit to the OMEG Institute at Soongsil University and the valuable discussions held during the 70th OMEG-SSANP Workshop 2023. 
A.J.A. gratefully acknowledges the support received from the Special Postdoctoral Researcher (SPDR) Program at RIKEN. 
The work of P.T.P.H. was supported by the National Research Foundation of Korea (NRF) grants funded by the Korean government (MSIT) Nos. 2018R1A5A1025563, 2022R1A2C1003964, and 2022K2A9A1A0609176, and by the PUTI Q1 Research Grant from the University of Indonesia (UI) under contract No. NKB-442/UN2.RST/HKP.05.00/2024. 
The work of K.T.~was supported by Conselho Nacional de Desenvolvimento Cient\'{i}fico e Tecnol\'ogico (CNPq, Brazil), Processes No.~313063/2018-4, No.~426150/2018-0, and No. 304199/2022-2, and FAPESP Process No.~2019/00763-0 and No.~2023/07313-6, and his work was also part of the projects, Instituto Nacional de Ci\^{e}ncia e Tecnologia - Nuclear Physics and Applications (INCT-FNA), Brazil, Process No.~464898/2014-5.

%----------------------------------------------------------------
\bibliography{reference}
%----------------------------------------------------------------
\end{document}